\documentclass[aps,pra,reprint,showpacs,floatfix,amsmath,amssymb,10pt]{revtex4-1} 

\usepackage[pdftex]{graphicx}       
\usepackage{braket}
\usepackage{epstopdf}
\usepackage{txfonts}        
\usepackage{mathtools}
\usepackage{mathrsfs}

\usepackage{hyperref} 
\hypersetup{colorlinks,citecolor=blue,filecolor=blue,linkcolor=blue,urlcolor=blue}

\usepackage{amsmath}				
\newcommand{\scE}{\mathcal{E}}
\newcommand{\scS}{\mathcal{S}}
\newcommand{\scY}{\mathcal{Y}}
\newcommand{\scP}{\mathcal{P}}

\newcommand{\aho}{a_\text{ho}}

\newcommand{\rc}{r_\text{c}}

\newcommand{\rr}{{\mathbf r}}	

\newcommand{\ww}{{\mathbf w}}
\newcommand{\uu}{{\mathbf u}}

\newcommand{\vdd}{V_\text{dd}(\rr)}
\newcommand{\cdd}{C_\text{dd}}
\newcommand{\xx}{{\mathbf x}}

\newcommand{\eref}[1]{Eq.~\eqref{#1}}
\newcommand{\fref}[1]{Fig.~\ref{#1}}

\begin{document}

\title{Analysis of two and three dipolar bosons in a spherical harmonic trap}

\author{C. J. Bradly} \email{cbradly@pgrad.unimelb.edu.au}
\author{H. M. Quiney}
\author{A. M. Martin}
\affiliation{School of Physics, University of Melbourne, Victoria 3010, Australia}

\date{\today}

\begin{abstract}
As dipolar gases become more readily accessible in experiment there is a need to develop a comprehensive theoretical framework of the few-body physics of these systems. 
Here, we extend the coupled-pair approach developed for the unitary two-component Fermi gas to a few-body system of dipolar bosons in a spherical harmonic trap. The long range and anisotropy of the dipole-dipole interaction is handled by a flexible and efficient correlated gaussian basis with stochastically variational optimization. Solutions of the two-body problem are used to calculate the eigenenergy spectrum and structural properties of three trapped bosonic dipoles. This demonstrates the efficiency and flexibility of the coupled-pair approach at dealing with systems with complex interactions.
\end{abstract}

\maketitle

\section{Introduction}

Advances in trapping of ultracold quantum gases have enabled the realization of systems with significant dipolar interactions. Experimentalists have been able to trap atoms that have a naturally large magnetic dipole moment, eg. ${}^{52}\text{Cr}$ \cite{Griesmaier2005,Bismut2010}, ${}^{168}\text{Er}$ \cite{Aikawa2012} and ${}^{164}\text{Dy}$ \cite{Lu2011,Lu2012}. These atoms have much stronger magnetic dipole moments compared with alkali atoms \cite{Lahaye2009}. Alternatively, clouds of polar molecules can be trapped and their electric dipole moment controlled by an external electric field \cite{Ni2008, Deiglmayr2008, Stuhl2012}. This interest in dipolar gases is due to the long range and anisotropy of the interaction between dipoles, which is in distinct contrast to the contact interaction \cite{Menotti2008}. Since the strength of the dipole-dipole interaction (DDI) is proportional to the square of the dipole moment, and combined with the use of Feshbach resonances to tune the van der Waals interaction {\em away} from unitarity, it is possible to create purely dipolar ultracold atomic gases \cite{Koch2008,Lahaye2009b}. This opens up a rich spectrum of new physics \cite{Micheli2006,Lahaye2009,Baranov2012} including dipolar Bose-Einstein condensates \cite{Griesmaier2005} and control of chemical reactions \cite{Ospelkaus2010,Quemener2012}. However, the dipolar interactions can also inhibit the ability to trap a stable cloud \cite{Lahaye2009}. 

 
Along with the experimental realization of ultracold dipolar gases the many-body dipolar gas has been studied theoretically \cite{Baranov2012,Lahaye2009}. Much work has also been done on the scattering theory of the DDI \cite{Deb2001,Gao1999,Kanjilal2008,Melezhik2003}, including the development of anisotropic pseudopotentials \cite{Derevianko2003}, prediction of Efimovian universal three-body bound states \cite{Wang2011,Wang2011b} and the scattering of dipoles in one- and two-dimensional geometries \cite{Volosniev2013,Ticknor2009,Ticknor2010,DIncao2011,Cai2010,Koval2014,Volosniev2011}. Reduced dimensional systems are particularly interesting for the appearance of dipolar confinement-induced resonances \cite{Giannakeas2013}. The dipolar two-body problem has also been studied in a spherically symmetric harmonic trapping potential \cite{Kanjilal2007,Wang2012}. However, the few-body physics of trapped dipolar atoms is not as advanced as for neutral atoms with a contact interaction. Here we extend techniques developed for the two-component Fermi gas \cite{Bradly2014} to a system of two and three three dipolar bosons in a spherically symmetric harmonic trap. This involves the application of a correlated gaussian basis optimized using the stochastic variational method \cite{Suzuki1998,Mitroy2013} that allows matrix elements to be calculated quickly and efficiently. We also employ the coupled-pair approach as a physically consistent way to counter the increase in complexity of exact diagonalization problems as the number of particles increases. The DDI potential allows for many resonances across multiple scattering channels and choosing the important correlations is a good test for extending these methods to more complex systems.







\section{Dipole-dipole scattering}
Before considering trapped particles we look at the low-energy scattering properties of two aligned dipoles with a view to its application to the harmonically trapped system. The DDI potential may be written
\begin{equation}
	\vdd   = \frac{\cdd}{4\pi} \frac{1-3\cos^2\theta}{r^3} = -\frac{\hbar^2}{M} \frac{D}{r^3} \sqrt{\frac{16\pi}{5}} Y_2^0(\hat{\rr})
	\label{eq:DDI}
\end{equation}
where $\theta$ is the angle to the axis of polarization (assumed to be the $z$-axis). In this work we choose the relative coordinates so that all reduced masses are the same as the single particle mass $M$. The strength of the interaction is determined by the type of particle. For magnetic dipoles with magnetic moment $\mathbf{d}_\text{m}$, \smash{$\cdd=\mu_0 d_\text{m}^2$}, while for molecules with polarizability $\alpha$ in an electric field $\mathbf{\mathcal{E}}$, \smash{$\cdd = \scE^2\alpha^2/\epsilon_0$}. In either case, it is useful to parameterize the strength of the interaction with an associated length scale, \smash{$D=M\cdd/4\pi\hbar^2$}.



The DDI 
is both long-ranged and anisotropic. The anisotropy means that two dipoles can experience attraction, if their collision axis is close to the $z$-axis, or repulsion, if it is closer to the transverse plane. 
Although we do not have global spherical symmetry, since the anisotropy is expressed as a single spherical harmonic the DDI neatly couples different partial waves. Furthermore, since the dipoles are aligned with the $z$-axis, the system retains axial symmetry. Therefore, we suppose that in each channel the asymptotic form of the wavefunction is \smash{$\psi_l(\rr) = r u_l(r)Y_l^{m}(\theta,\phi) $} and the phase shifts are found by integrating the radial Schr\"odinger equation with the potential of \eref{eq:DDI}, and matching the solutions to spherical waves at large separation distance. This obtains a set of coupled radial equations
\begin{equation}
	\frac{d^2u_l}{dr^2}+\left( k^2 - \frac{l(l+1)}{r^2} \right)u_l + \frac{2D}{r^3} \sum_{l'} \scY(m;l',l)u_{l'} = 0,
	\label{eq:DipolarSE}
\end{equation}
where the angular integration has been performed to give the coupling coefficients \cite{Wang2012}
\begin{alignat}{1}
	\scY(m;l',l) &= \sqrt{\frac{16\pi}{5}}\int\!\! d\hat\rr {Y_{l'}^m}^*(\hat\rr)Y_2^0(\hat\rr)Y_l^m(\hat\rr)  \nonumber \\
				&= (-1)^m 2 \sqrt{(2l'+1)(2l+1)} \begin{pmatrix} l' & 2 & l \\ 0&0&0 \end{pmatrix}	\begin{pmatrix} l' & 2 & l  \\ -m&0&m \end{pmatrix},
	\label{eq:ThreeYSymbol}
\end{alignat}
expressed by Wigner 3-$j$ symbols. These coefficients reveal that the dipole interaction only allows coupling between channels for which \smash{$l-l' = -2,0,2$}. Since the DDI preserves the axial symmetry of the system different $m$ channels decouple, although they are not necessarily degenerate. However, since higher $m$ states cannot couple to lower $l$ states (if \smash{$|m|>l$}), we expect that higher $m$ channels will have little impact on the bosonic ground state and here we restrict our study to \smash{$m=0$}. The partial wave expansion not only handles the anisotropy of the DDI, but also determines the parity of the wavefunction and thus its symmetry. Aligned bosonic dipoles only admit states with even angular momentum $l$, thus neatly matching the allowed $l$ couplings of the DDI. 

Having dealt with the anisotropy, we now turn to the long-ranged nature of the potential. Ignoring for the moment the coupling to other channels, we must deal with an attractive $-1/r^3$ potential in each channel, since the coupling coefficients $\scY(m;l',l)$ are non-negative for \smash{$m=0$}. Recently, some advanced methods have been developed including a multichannel pseudopotential for dipolar scattering \cite{Derevianko2003} and exact solutions for the isotropic pure-repulsive $1/r^3$ potential based on quantum defect theory \cite{Gao1999}. However, in view of the application of the coupled-pair approach to a system of trapped dipoles we require a more straightforward approach and replace $\vdd$ with \smash{$\tanh^{15}(r/\rc)\vdd$} \cite{Wang2012}, where $\rc$ is the cutoff distance. The cutoff avoids the unphysical $1/r^3$ potential at short distances and since it is smooth it will be better handled by the correlated gaussian basis, compared to e.g.~a hard-sphere cutoff \cite{Kanjilal2007}. Note that, apart from being large enough to make the cutoff steep, the power of 15 is arbitrary and the smoothness of $\tanh^{15}(r/\rc)$ is more important.

To solve \eref{eq:DipolarSE} we employ the Johnson log-derivative algorithm \cite{Johnson1973}. The algorithm is applied starting from the inner boundary \smash{$r=0$} out to some large $r_\text{max}$, at a fixed small collision energy \smash{$k=\sqrt{2\times 9.36\times 10^{-9}}/\rc$}. This is much the same as a single channel problem but since the DDI is long-ranged, we must take extra care to ensure that $r_\text{max}$ is sufficiently large. For our purposes, the short-range cut-off $\rc$ serves as a convenient length scale and the strength of the DDI is parameterized by the ratio $D/\rc$. Once the inner-region solution is found it is matched to the solution in the outer region at $r_\text{max}$. Although a log-derivative matrix is slightly more complicated, this outer solution can still be expressed in terms of Bessel functions, similar to single channel scattering. The Johnson log-derivative method then neatly allows for the determination of the $K$-matrix from the solution at  $r_\text{max}$. The elements of the $K$-matrix encode the phase shifts, which in turn can be expressed as scattering lengths $a_{ll'}$, where the channels are labeled by their angular momenta.

\begin{figure}
	\centering
	\includegraphics[width=\columnwidth]{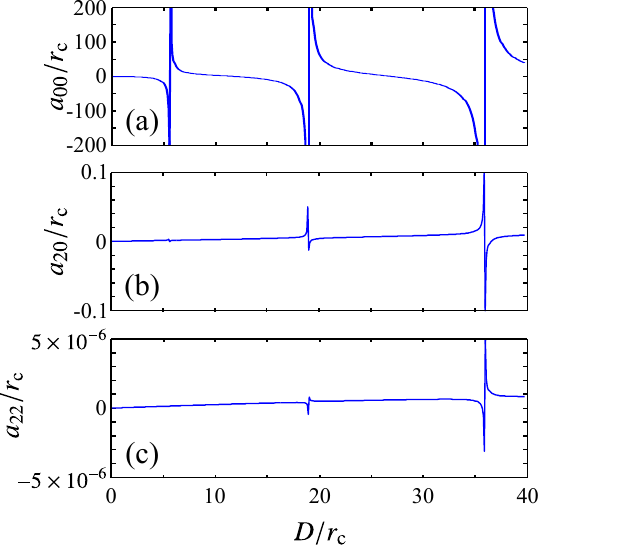}
	\caption{Scattering lengths (a) $a_{00}/\rc$, (b) $a_{20}/\rc$ and (c) $a_{22}/\rc$ for two bosons with DDI showing the first few resonances as a function of dipole length $D$.}
	\vspace{-0.5cm}
	\label{fig:DDIScattering}
\end{figure}

In \fref{fig:DDIScattering} we plot the scattering lengths (a) $a_{00}/\rc$, (b) $a_{20}/\rc$ and (c) $a_{22}/\rc$ for two bosonic dipoles. There is a sequence of resonances that appear over a large range of $D/\rc$, beginning with \smash{$D/\rc=5.86,19.28,36.34$}. The resonances are broadest in the $s$-wave channel (a) but appear at the same interaction strengths for higher partial waves. We can also see in (b) and (c) that away from resonances the scattering lengths change linearly with $D/\rc$. This linear dependence in the higher partial waves is predicted by the Born approximation for the DDI \cite{Kanjilal2008}. However, it also predicts \smash{$a_{00}/\rc=0$}, which is clearly not the case. The behavior of $a_{00}$, including the width and position of the resonances is affected by the short-range details of the potential, in particular, the cut-off distance $\rc$ as well as the type of cut-off function. It is not expected that the dipolar gas will display universal properties with respect to the short-ranged physics and we have picked the $\tanh^{15}(r/\rc)$ cut-off with the gaussian basis in mind.

\section{Trapped dipolar bosons}

Having mapped out the scattering properties of the DDI, we now turn to the problem of dipolar atoms in a spherically symmetric harmonic trapping potential. 
The focus here is to test the application of a correlated gaussian basis, and in particular the coupled pair approach, to a system with a more complicated interaction potential. Thus, we will only look at \smash{$N=2$} and \smash{$N=3$}. The relative Hamiltonian for $N$ trapped dipolar atoms is
\begin{alignat}{1} 
	H &= \sum_{i=1}^{N-1}\left[-\frac{\hbar^2}{2M}\nabla_i^2 + \frac{1}{2}M\omega^2 x_i^2 \right] + \sum_{i<j}^{} V_\text{dd}(\ww_{ij}^\intercal\xx),
	\label{eq:3bHamiltonian}
\end{alignat}
where the first term is the kinetic energy and the second is the external spherically symmetric trapping potential of frequency $\omega$. The coordinates are contained in $\xx$, which is a supervector of all relative coordinate vectors. The coordinates are similar to Jacobi coordinates but rescaled such that all reduced masses are equal to the single-particle mass $M$. The particular coordinates for each $N$ are detailed later. The last term in \eref{eq:3bHamiltonian} contains the DDI potentials between all possible pairs of particles. The selection vector $\ww_{ij}$, defined by $\ww_{ij}^\intercal\xx=\rr_i-\rr_j$, picks out the relative vector $\rr_i-\rr_j$ from the full set of relative coordinates $\xx$.



The harmonic trapping potential promotes the use of a correlated gaussian basis for the radial degrees of freedom and the long-ranged nature of the DDI is a good test for the flexibility and efficiency of this basis. A gaussian basis is also effective for the short-ranged correlations in the system provided we employ a smooth cut-off function, as opposed to a hard-sphere boundary condition. The anisotropy of the DDI means that the basis must also incorporate angular degrees of freedom and, like the free-space scattering, this is done by having basis elements with well-defined total relative angular momentum $\{l,m\}$. The total (unnormalized and unsymmetrized) basis element is therefore
\begin{equation}
	\braket{\xx|A,\uu;l} = \exp\left( -\tfrac{1}{2}\xx^\intercal A\, \xx \right)| \uu^\intercal\xx|^l Y_l^m\left( \widehat{\uu^\intercal\xx} \right),
	\label{eq:GVBasisFunction}
\end{equation}
where $l$ is the total relative angular momentum, $A$ is an $(N-1)\times(N-1)$ diagonal matrix whose $i^\text{th}$ entry is $1/\alpha_i^2$, where $\alpha_i$ is the Gaussian widths for the $i^\text{th}$ coordinate and $\uu$ is a normalized global vector with \smash{$N-1$} components that describes the distribution of internal angular momentum among the angular degrees of freedom. The gaussian widths $\alpha_i$ and the components of $\uu$ are the \smash{$2N-3$} nonlinear variational parameters of the basis. Since angular momentum is not conserved in the presence of the DDI, it is preferable to use a continuous variational parameter for describing the angular degrees of freedom, as opposed to coupled sets of discrete angular momenta. All relevant matrix elements with this basis can be calculated analytically \cite{Suzuki1998}. We note that, even though the evaluation of these matrix elements is efficient, the inclusion of angular momentum and degrees of freedom does require more operations than for the isotropic gaussian-only basis used for the two-component Fermi system \cite{Bradly2014}. Furthermore, this basis can only describe states with natural parity, precluding the investigation of three fermionic trapped dipoles without significant extension.

We first consider \smash{$N=2$} trapped dipolar bosons, since the \smash{$N=3$} system is built from these solutions. The DDI will couple states with different angular momentum, but since there is only a single relative coordinate, \smash{$\xx=(\rr_1-\rr_2)/\sqrt{2}$}, we do not need the global vector. In fact, the angular degrees of freedom manifest as a standard spherical harmonic in each basis element. The machinery of the global vector representation is simplified and is essentially equivalent to use of the coefficients \smash{$\scY(m;l',l)$} of \eref{eq:ThreeYSymbol}. Therefore, for \smash{$N=2$}, the angular degrees of freedom are determined by the number of angular momentum channels included. Like the free-space scattering case, bosons are distinguished by restricting the angular momentum channels to even  $l$.

While the channel coupling is simplified, the two-body problem is nonetheless a test of using a gaussian basis to tackle a system with long-range interactions. We apply the stochastic variational method (SVM) \cite{Suzuki1998} to choose the gaussian widths, but with some extra considerations as to the allowed range they can take. For the trapped system, $\aho$ is taken to be the standard length-scale, but we now also have $D$, characterizing the long-ranged strength of the DDI, and $\rc$, characterizing the short-range cut-off. Since the form of the DDI in \eref{eq:DDI} is only valid at sufficiently large separations, we do not want to examine the zero-ranged limit \smash{$\rc\to 0$}. Instead, we choose \smash{$\rc/\aho=0.01$} (i.e.~maintaining the limit \smash{$\rc\ll\aho$}) and examine changing $D/\aho$. With $\rc$ given, we can use the scattering results to consistently determine the strength of the DDI. At the other end of the scale, the long-ranged nature of the interaction means that $\aho$ is not as hard an upper limit as for the unitary two-component Fermi system \cite{Bradly2014}, where the SVM would optimize the gaussian widths to \smash{$\alpha_{i}/\aho\lesssim 1.1$}, reflecting that $\aho$ was the largest physical correlation in the problem. For dipolar particles, we still insist that \smash{$\alpha_{i}\gg\aho$} would violate the idea that the particles have low energy, but even for small $D/\aho$ it is necessary to allow the widths up to \smash{$\alpha_{i}/\aho\approx 3$}. Even if correlations larger than the size of the trap contribute only a small amount to the problem, it is physically meaningful to include them due to the long range of the DDI. These bounds are used to semi-stochastically select the trials in the SVM, meaning that while most trials satisfy \smash{$\rc<\alpha_i<3\aho$}, 10\% of trials satisfy \smash{$\rc>\alpha_i$} and 10\% satisfy \smash{$3\aho<\alpha_i$}. This ensures that both long- and short-ranged correlations are probed, subject to optimization.

\begin{figure}[t!]
	\centering
	\includegraphics[width=\columnwidth]{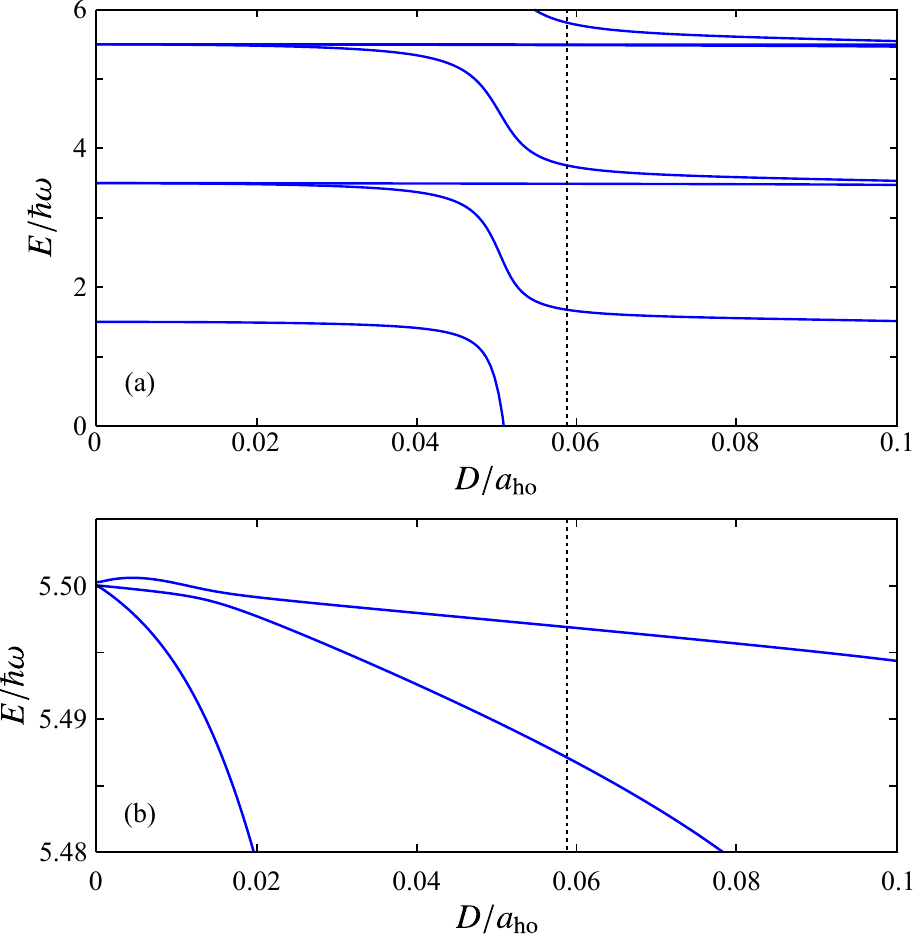}
	\caption{(a) The energy spectrum of \smash{$N=2$} bosonic dipoles  near the first resonance. The short-range cut-off is fixed at \smash{$\rc/\aho=0.01$} so the resonance occurs at \smash{$D/\aho=0.0586$}, as shown by the dotted vertical line. (b) A close-up near \smash{$E=5.5\hbar\omega$}, which shows one state decreasing much faster than the others. These states can be characterized by their angular-momentum in the non-interacting limit, which are, from bottom to top, \smash{$l=0,2,4$}, respectively. Note also the avoided crossing between the other two states at small $D/\aho$, which is due to the short-range cut-off imposed on the DDI.}%
	\vspace{-0.5cm}
	\label{fig:2bDDISpectrum}%
\end{figure}

The energy spectrum is calculated by solving a generalized eigenvalue problem for a range of the DDI strength, $D$. The correlated gaussian basis allows fast and efficient calculation of the matrix elements of the Hamiltonian, \eref{eq:3bHamiltonian}, and the overlaps between basis elements. In \fref{fig:2bDDISpectrum}(a) we plot the energy spectrum for \smash{$N=2$} trapped bosonic dipoles as a function of the interaction strength $D/\aho$ near the first resonance. The basis includes the first four angular momentum channels, \smash{$l=0,2,4,6$} and uses a fixed short-range cut-off parameter \smash{$\rc/\aho=0.01$}. On this scale, the position of the first resonance occurs at \smash{$D/\aho=0.0586$}, and is marked by a vertical dotted line. Not all states are strongly affected by the DDI resonance, with only one state in each level `stepping down' by $2\hbar\omega$ across the resonance. If we characterize the states by their angular momentum in the non-interacting limit (\smash{$D/\aho=0$}) then it is the \smash{$l=0$} state which is most affected. Away from resonance, the DDI has only a weak effect, which is hard to see on this scale but is visible in \fref{fig:2bDDISpectrum}(b), which shows a close-up of the spectrum near \smash{$E=5.5\hbar\omega$}. On this scale we can see that these energies decrease as the DDI strength increases. Note also the avoided crossing between the higher states at small $D/\aho$ that is due to the short-range cut-off imposed on the DDI.
Recently, Refs.~\cite{Giannakeas2013} have shown that confinement-induced resonances can occur in dipolar systems that are confined by a quasi-1D external trapping potential. However, the resonance that we study here is not confinement-induced but is the resonance from the DDI only.

In principle, the SVM optimization should be performed separately at each value of $D/\aho$. In practice however, we find that for a sufficiently large basis (more than 30 gaussians in each angular momentum channel), refining the SVM optimization does not obtain a different basis as $D/\aho$ is varied. That is, provided that the basis contains terms at all relevant length scales then it can efficiently handle the weakly interacting and resonant regimes equally.

\section{Results for $N=3$}
For \smash{$N=3$} dipolar bosons, we employ the coupled-pair approach \cite{Bradly2014} with the global vector representation. The relative coordinates are \smash{$\xx_1=(\rr_1-\rr_2)/\sqrt{2}$} and \smash{$\xx_2=\sqrt{2/3}[(\rr_1+\rr_2)/2-\rr_3]$}. This means we allow for the explicit interaction between particles 1 and 2, but not between this pair and particle 3. That is, the scattering channel for \smash{$N=3$} has one interacting pair correlation (IPC), where the range of gaussian widths must account for the DDI as discussed above, and one noninteracting correlation (NIC), where the range of gaussian widths only needs to account for the noninteracting length scales, chiefly $\aho$. Within these ranges, the gaussian widths are optimized stochastically as two independent two-body problems and then combined to calculate the matrix elements for the \smash{$N=3$} system. Other scattering channels are accounted for by symmetrizing the basis by application of \smash{$\scS = 1 + \scP_{13} + \scP_{23}$} to the basis elements. As with the free-space and \smash{$N=2$} cases, bosons require even total relative angular momentum to maintain the correct symmetry and this means we do not need to explicitly apply $\scP_{13}$. Stochastic variation of the global vector $\uu$ does the work in choosing the internal angular momenta.

\begin{figure}[t!]
	\centering
	\includegraphics[width=\columnwidth]{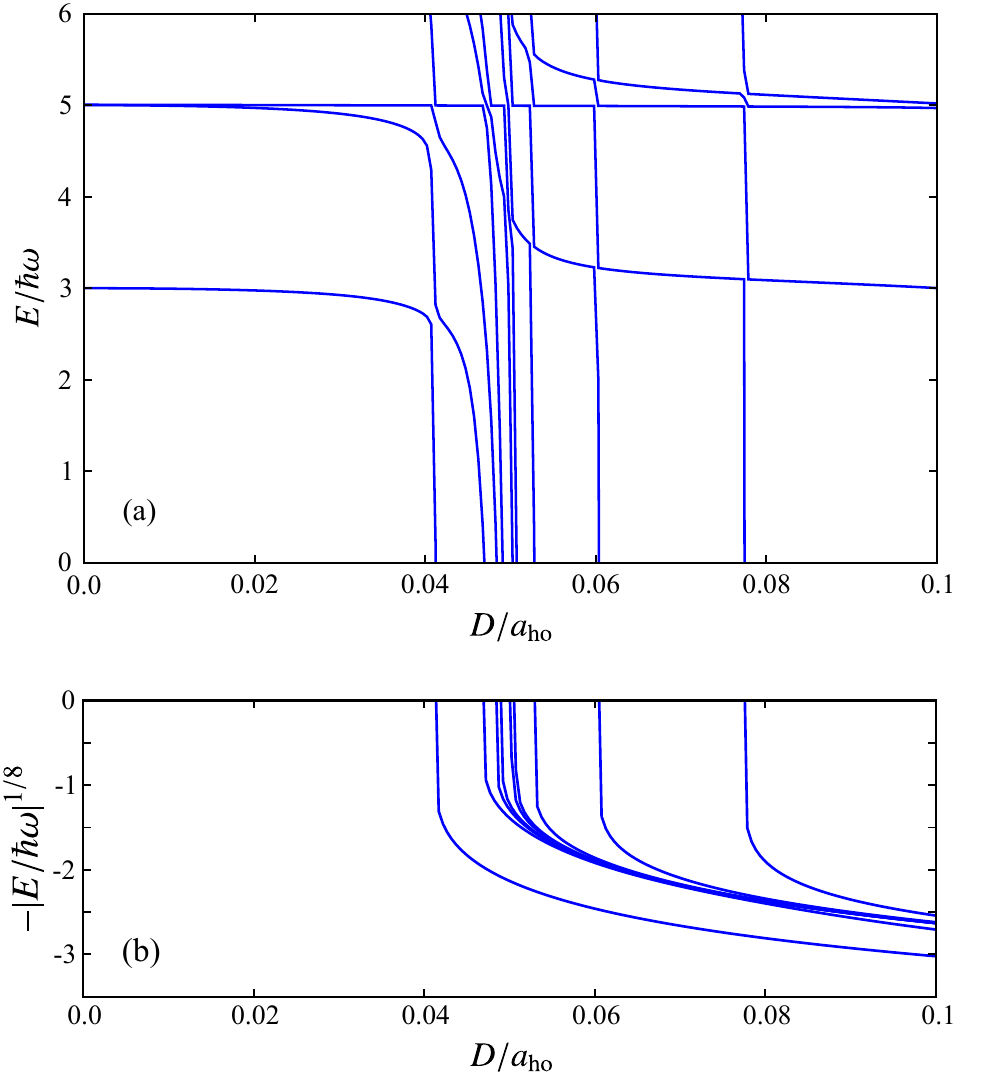}
	\caption{(a) The energy spectrum of \smash{$N=3$} bosonic dipoles for near the first resonance at fixed short-range cut-off \smash{$\rc/\aho=0.01$}. The basis includes the first four angular momentum channels. Many bound states appear to cluster around the position of the two-body DDI resonance at \smash{$D/\aho=0.0586$}. The ground state diverges before this point differently to the others as shown in (b).}
	\vspace{-0.5cm}
	\label{fig:3bDDISpectrum}
\end{figure}

With these alterations for bosons and the DDI, we apply the coupled-pair approach and the SVM as before. Each two-body subsystem is solved independently using SVM to determine the gaussian widths. These are combined to make the correlation matrices $A$ for the three-body problem. The gaussian part of the basis is then replicated for each value of $l$. Again, with a sufficiently large basis, it is not necessary to repeat this procedure for different values of $D/\aho$ since the refinement phase of the SVM does not improve upon the basis as $D/\aho$ in changed by a small amount. Upon solving the generalized eigenvalue problem, \fref{fig:3bDDISpectrum} shows the energy spectrum for three bosonic dipoles. Here, the basis includes the first four angular momentum channels, \smash{$l=0,2,4,6$} with \smash{$\uu=(1,0)$} (as justified below). There are many more bound states than for \smash{$N=2$}, but they mostly cluster around the two-body resonance at \smash{$D/\aho=0.0586$}. These states are due to more particles being able to interact. However, the ground state diverges at a smaller value of $D/\aho$ and behaves differently to the others. This is best seen in \fref{fig:3bDDISpectrum}(b), where the bound state energies are plotted on a $|E|^{1/8}$ scale. Here we see that most of the bound states converge in the deeply bound limit but the ground state displays distinctly different behavior and stays separate. 
We note also that although variational optimization strictly applies to only the ground state, the \smash{$N=3$} problem is small enough that a large basis can be used, and the energies of excited states are converged within the line width of \fref{fig:3bDDISpectrum}.


The optimization of the global vector $\uu$ is more complicated. The spectrum presented in \fref{fig:3bDDISpectrum} used \smash{$\uu=(1,0)$} for all basis states, which requires some justification. For \smash{$N=3$} the global vector can be written as \smash{$\uu = (\cos \varphi,\sin\varphi)$}; essentially reducing to the single variational parameter $\varphi$, for which it is sufficient to consider \smash{$0\le\varphi\le \pi/2$}. Since \smash{$\uu$} does not appear for \smash{$N=2$}, the coupled-pair approach does not apply to this variational parameter. Instead, we must investigate the angular degrees of freedom separately at the three-body level. Since \smash{$\uu$} does not appear for \smash{$N=2$}, the coupled-pair approach does not apply to this variational parameter. Instead, we must investigate the angular degrees of freedom separately at the three-body level. We still wish to avoid the labor of applying SVM to the entire problem so, similar to the scan over interaction strength $D$, we apply the coupled-pair SVM approach once with a large basis and then reuse the gaussian widths when varying $\varphi$. We note that for a large enough basis, such that the radial part has converged (in this work, at least 150 gaussians were used per angular momentum channel), the result is independent of the initial value of $\varphi$ used. It is also not necessary to optimize $\varphi$ stochastically because the coupled-pair approach has decoupled the variational parameter $\varphi$ from the gaussian widths $\alpha_i$.

\begin{figure}[t!]
	\centering
	\includegraphics[width=\columnwidth]{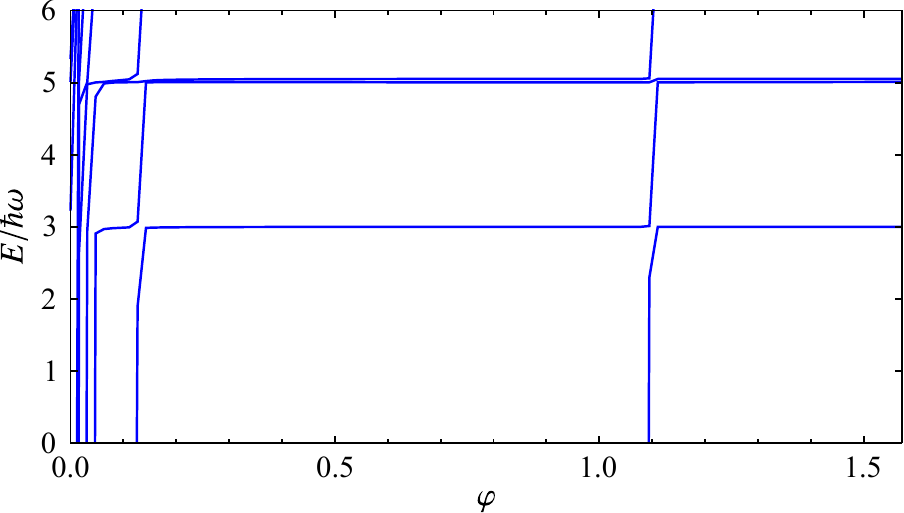}
	\caption{The \smash{$N=3$} bosonic dipolar spectrum for varying global vector parameter $\varphi$. The strength of the DDI is fixed just above the resonance, at \smash{$D/\aho=0.06$}. In this picture, more bound states (i.e.~lower energies) appear at \smash{$\varphi=0$}, meaning that this is the optimal value of the variational parameter.}
	\vspace{-0.5cm}
	\label{fig:3bVaryTheta}%
\end{figure}

In \fref{fig:3bVaryTheta} we plot the spectrum at \smash{$D/\aho=0.06$}, just above the resonance, for \smash{$0\le\varphi\le \pi/2$}. At \smash{$\varphi=\pi/2$} there are no bound states, implying the resonance appears at a higher \smash{$D/\aho$}. Bound states appear as $\varphi$ is lowered, meaning that the position of the resonance is lowered. This continues until \smash{$\varphi=0$} for which the energies are are at their lowest value. In accordance with the variational principle, we should therefore take \smash{$\varphi=0$} as the optimal value. Since $\varphi$ is a smooth variational parameter, it is not generally clear how it determines the internal distribution of angular momentum. However, for extremal values it implies that all the angular momentum of a state is in one two-body sub-system. For example, \smash{$\varphi=0$}, i.e.~\smash{$\uu=(1,0)$}, means that the total relative angular momentum $l$ of a basis state is entirely found in the correlation of particles 1 and 2, while the motion of particle 3 relative to the pair is $s$-wave. Although this effect is best visualized at an interaction strength just above the resonance, we note that \smash{$\varphi=0$} gives the lowest energy for all $D/\aho$ considered in this work.

\begin{figure}[t!]
	\centering
	\includegraphics[width=\columnwidth]{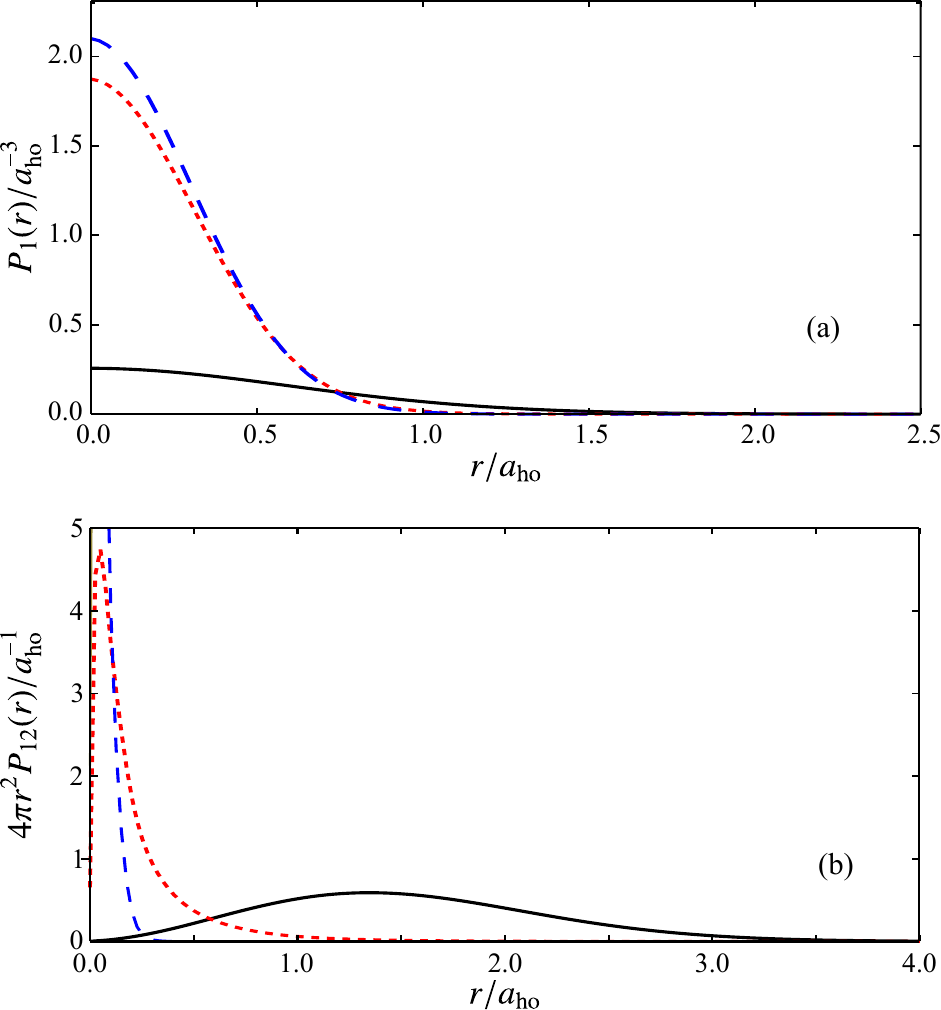}
	\caption{(a) The single-particle density, $P_1(r)$ and (b) the pair-correlation function, $P_{12}(r)$, for \smash{$N=3$} aligned dipolar bosons. The three curves are for \smash{$D/\aho=0.05$} (solid black), \smash{$D/\aho=0.0586$} (dotted red) and \smash{$D/\aho=0.06$} (dashed blue).}
	\vspace{-0.5cm}
	\label{fig:DDIStructural}
\end{figure}

Finally, we can also look at the structural properties for \smash{$N=3$} trapped dipolar bosons. By diagonalizing the relative Hamiltonian we not only obtain the energy spectrum but the relative wavefunction. Combining this with the center-of-mass wavefunction we obtain the total wavefunction $\Psi(\xx)$.
From $\Psi(\xx)$ we can calculate a general structural property
\begin{alignat}{1}
	P(r)=\int\!\!  d\rr'\,\frac{\delta(r-r')}{4\pi r'^2} \int\!\! d\xx\, \delta(\ww^\intercal\xx-\rr') |\Psi(\xx)|^2,
	\label{eq:GeneralDensity}
\end{alignat}
where $\rr$ (and $\rr'$) is a coordinate describing the property of interest and $P(r)$ is normalized to unity. Here $\xx$ is a general set of coordinates such as the center-of-mass plus relative coordinates as defined above or the single-particle coordinates and $\ww$ is a selection vector that picks out $\rr$ from $\xx$, much like for the interaction potential. In particular we calculate the single-particle reduced density \smash{$P_1(r)/\aho^{-3}$}, with \smash{$\rr=\rr_1$} in \eref{eq:GeneralDensity} and the (scaled) pair correlation function \smash{$4\pi r^2 P_{12}(r)/\aho^{-1}$}, with \smash{$\rr=\rr_1-\rr_2$} in \eref{eq:GeneralDensity}. 
The wavefunction for dipolar bosons includes contributions from higher angular momentum channels, but since we are only looking at the \smash{$m=0$} subspace, both the single-particle density, $P_1(r)$, and pair correlation function, $P_{12}(r)$, remain functions of $r$ only.

In \fref{fig:DDIStructural}(a) we plot the single particle density function, $P_1(r)$, for different values of $D/\rc$. Below the resonance, \smash{$D/\aho=0.05$} (solid black), the density is a low wide peak extending over the whole trap. This reflects the regime of weak interactions. At the first DDI resonance, \smash{$D/\aho=0.0586$} (dotted red), the single-particle density is more sharply peaked with an exponential decay beyond \smash{$r=\aho$}. Just above the resonance, \smash{$D/\aho=0.06$} (dashed blue), the density is more sharply peaked still. This shows that the atoms are being held closer together as the DDI strength increases and becomes more attractive. The scaled pair-correlation function, $4\pi r^2P_{12}(r)$, is shown in \fref{fig:DDIStructural}(b) for the same values of $D/\aho$. Below resonance (solid black) the pair-correlation function has a single peak near \smash{$r/\aho\approx1.5$}, suggesting that without strong interactions the particles are dispersed. Near resonance (dotted red) the correlation is peaked near \smash{$r/\aho\approx0$}. In the limit $r/\aho\to 0$, $4\pi r^2P_{12}(r)$ would go to a finite value except that we have a short-ranged cut-off. This is similar to the near-vertical drop seen for the two-component unitary Fermi gas \cite{Bradly2014}. However, unlike the Fermi gas, there is no second peak, which means that the atoms are not weakly bound. Above resonance (dashed blue) the scaled correlation function diverges as \smash{$r/\aho\to 0$} and decays to zero much quicker than near resonance, which reflects that above resonance the ground state is in a deeply bound regime, as expected from the energy spectrum of \fref{fig:3bDDISpectrum}.

\section{Conclusion}

The goal of this work was to apply the coupled-pair approach to a system with a more complicated interparticle interaction. The dipole-dipole interaction is both long-ranged and anisotropic and we have considered the problem of \smash{$N=2$} and \smash{$N=3$} bosonic dipoles in a harmonic trap by extending the gaussian basis to include angular degrees of freedom. We have investigated the scattering properties of the dipole-dipole interaction and then calculated the eigenenergy spectrum and structural properties of \smash{$N=3$} aligned dipoles near a resonance. Like the Fermi gas, deeply bound states appear near the resonance but here they form in all angular momentum channels. The ground state is distinctly differently to the excited states in the deeply-bound regime. The structural properties, especially the pair-correlation function, show that the DDI becomes more attractive as its strength is increased, and across the resonance the atoms form more deeply bound pairs, while on resonance the ground state has similarities to the unitary Fermi gas. This demonstrates that the coupled-pair approach can be applied to more complex systems with long-ranged and anisotropic interactions.




\bibliography{bib_dipolar}

\begin{thebibliography}{37}%
\makeatletter
\providecommand \@ifxundefined [1]{%
 \@ifx{#1\undefined}
}%
\providecommand \@ifnum [1]{%
 \ifnum #1\expandafter \@firstoftwo
 \else \expandafter \@secondoftwo
 \fi
}%
\providecommand \@ifx [1]{%
 \ifx #1\expandafter \@firstoftwo
 \else \expandafter \@secondoftwo
 \fi
}%
\providecommand \natexlab [1]{#1}%
\providecommand \enquote  [1]{``#1''}%
\providecommand \bibnamefont  [1]{#1}%
\providecommand \bibfnamefont [1]{#1}%
\providecommand \citenamefont [1]{#1}%
\providecommand \href@noop [0]{\@secondoftwo}%
\providecommand \href [0]{\begingroup \@sanitize@url \@href}%
\providecommand \@href[1]{\@@startlink{#1}\@@href}%
\providecommand \@@href[1]{\endgroup#1\@@endlink}%
\providecommand \@sanitize@url [0]{\catcode `\\12\catcode `\$12\catcode
  `\&12\catcode `\#12\catcode `\^12\catcode `\_12\catcode `\%12\relax}%
\providecommand \@@startlink[1]{}%
\providecommand \@@endlink[0]{}%
\providecommand \url  [0]{\begingroup\@sanitize@url \@url }%
\providecommand \@url [1]{\endgroup\@href {#1}{\urlprefix }}%
\providecommand \urlprefix  [0]{URL }%
\providecommand \Eprint [0]{\href }%
\providecommand \doibase [0]{http://dx.doi.org/}%
\providecommand \selectlanguage [0]{\@gobble}%
\providecommand \bibinfo  [0]{\@secondoftwo}%
\providecommand \bibfield  [0]{\@secondoftwo}%
\providecommand \translation [1]{[#1]}%
\providecommand \BibitemOpen [0]{}%
\providecommand \bibitemStop [0]{}%
\providecommand \bibitemNoStop [0]{.\EOS\space}%
\providecommand \EOS [0]{\spacefactor3000\relax}%
\providecommand \BibitemShut  [1]{\csname bibitem#1\endcsname}%
\let\auto@bib@innerbib\@empty
\bibitem [{\citenamefont {Griesmaier}\ \emph {et~al.}(2005)\citenamefont
  {Griesmaier}, \citenamefont {Werner}, \citenamefont {Hensler}, \citenamefont
  {Stuhler},\ and\ \citenamefont {Pfau}}]{Griesmaier2005}%
  \BibitemOpen
  \bibfield  {author} {\bibinfo {author} {\bibfnamefont {A.}~\bibnamefont
  {Griesmaier}}, \bibinfo {author} {\bibfnamefont {J.}~\bibnamefont {Werner}},
  \bibinfo {author} {\bibfnamefont {S.}~\bibnamefont {Hensler}}, \bibinfo
  {author} {\bibfnamefont {J.}~\bibnamefont {Stuhler}}, \ and\ \bibinfo
  {author} {\bibfnamefont {T.}~\bibnamefont {Pfau}},\ }\href {\doibase
  10.1103/PhysRevLett.94.160401} {\bibfield  {journal} {\bibinfo  {journal}
  {Phys. Rev. Lett.}\ }\textbf {\bibinfo {volume} {94}},\ \bibinfo {pages}
  {160401} (\bibinfo {year} {2005})}\BibitemShut {NoStop}%
\bibitem [{\citenamefont {Bismut}\ \emph {et~al.}(2010)\citenamefont {Bismut},
  \citenamefont {Pasquiou}, \citenamefont {Mar\'echal}, \citenamefont {Pedri},
  \citenamefont {Vernac}, \citenamefont {Gorceix},\ and\ \citenamefont
  {Laburthe-Tolra}}]{Bismut2010}%
  \BibitemOpen
  \bibfield  {author} {\bibinfo {author} {\bibfnamefont {G.}~\bibnamefont
  {Bismut}}, \bibinfo {author} {\bibfnamefont {B.}~\bibnamefont {Pasquiou}},
  \bibinfo {author} {\bibfnamefont {E.}~\bibnamefont {Mar\'echal}}, \bibinfo
  {author} {\bibfnamefont {P.}~\bibnamefont {Pedri}}, \bibinfo {author}
  {\bibfnamefont {L.}~\bibnamefont {Vernac}}, \bibinfo {author} {\bibfnamefont
  {O.}~\bibnamefont {Gorceix}}, \ and\ \bibinfo {author} {\bibfnamefont
  {B.}~\bibnamefont {Laburthe-Tolra}},\ }\href {\doibase
  10.1103/PhysRevLett.105.040404} {\bibfield  {journal} {\bibinfo  {journal}
  {Phys. Rev. Lett.}\ }\textbf {\bibinfo {volume} {105}},\ \bibinfo {pages}
  {040404} (\bibinfo {year} {2010})}\BibitemShut {NoStop}%
\bibitem [{\citenamefont {Aikawa}\ \emph {et~al.}(2012)\citenamefont {Aikawa},
  \citenamefont {Frisch}, \citenamefont {Mark}, \citenamefont {Baier},
  \citenamefont {Rietzler}, \citenamefont {Grimm},\ and\ \citenamefont
  {Ferlaino}}]{Aikawa2012}%
  \BibitemOpen
  \bibfield  {author} {\bibinfo {author} {\bibfnamefont {K.}~\bibnamefont
  {Aikawa}}, \bibinfo {author} {\bibfnamefont {A.}~\bibnamefont {Frisch}},
  \bibinfo {author} {\bibfnamefont {M.}~\bibnamefont {Mark}}, \bibinfo {author}
  {\bibfnamefont {S.}~\bibnamefont {Baier}}, \bibinfo {author} {\bibfnamefont
  {A.}~\bibnamefont {Rietzler}}, \bibinfo {author} {\bibfnamefont
  {R.}~\bibnamefont {Grimm}}, \ and\ \bibinfo {author} {\bibfnamefont
  {F.}~\bibnamefont {Ferlaino}},\ }\href {\doibase
  10.1103/PhysRevLett.108.210401} {\bibfield  {journal} {\bibinfo  {journal}
  {Phys. Rev. Lett.}\ }\textbf {\bibinfo {volume} {108}},\ \bibinfo {pages}
  {210401} (\bibinfo {year} {2012})}\BibitemShut {NoStop}%
\bibitem [{\citenamefont {Lu}\ \emph {et~al.}(2011)\citenamefont {Lu},
  \citenamefont {Burdick}, \citenamefont {Youn},\ and\ \citenamefont
  {Lev}}]{Lu2011}%
  \BibitemOpen
  \bibfield  {author} {\bibinfo {author} {\bibfnamefont {M.}~\bibnamefont
  {Lu}}, \bibinfo {author} {\bibfnamefont {N.~Q.}\ \bibnamefont {Burdick}},
  \bibinfo {author} {\bibfnamefont {S.~H.}\ \bibnamefont {Youn}}, \ and\
  \bibinfo {author} {\bibfnamefont {B.~L.}\ \bibnamefont {Lev}},\ }\href
  {\doibase 10.1103/PhysRevLett.107.190401} {\bibfield  {journal} {\bibinfo
  {journal} {Phys. Rev. Lett.}\ }\textbf {\bibinfo {volume} {107}},\ \bibinfo
  {pages} {190401} (\bibinfo {year} {2011})}\BibitemShut {NoStop}%
\bibitem [{\citenamefont {Lu}\ \emph {et~al.}(2012)\citenamefont {Lu},
  \citenamefont {Burdick},\ and\ \citenamefont {Lev}}]{Lu2012}%
  \BibitemOpen
  \bibfield  {author} {\bibinfo {author} {\bibfnamefont {M.}~\bibnamefont
  {Lu}}, \bibinfo {author} {\bibfnamefont {N.~Q.}\ \bibnamefont {Burdick}}, \
  and\ \bibinfo {author} {\bibfnamefont {B.~L.}\ \bibnamefont {Lev}},\ }\href
  {\doibase 10.1103/PhysRevLett.108.215301} {\bibfield  {journal} {\bibinfo
  {journal} {Phys. Rev. Lett.}\ }\textbf {\bibinfo {volume} {108}},\ \bibinfo
  {pages} {215301} (\bibinfo {year} {2012})}\BibitemShut {NoStop}%
\bibitem [{\citenamefont {Lahaye}\ \emph
  {et~al.}(2009{\natexlab{a}})\citenamefont {Lahaye}, \citenamefont {Menotti},
  \citenamefont {Santos}, \citenamefont {Lewenstein},\ and\ \citenamefont
  {Pfau}}]{Lahaye2009}%
  \BibitemOpen
  \bibfield  {author} {\bibinfo {author} {\bibfnamefont {T.}~\bibnamefont
  {Lahaye}}, \bibinfo {author} {\bibfnamefont {C.}~\bibnamefont {Menotti}},
  \bibinfo {author} {\bibfnamefont {L.}~\bibnamefont {Santos}}, \bibinfo
  {author} {\bibfnamefont {M.}~\bibnamefont {Lewenstein}}, \ and\ \bibinfo
  {author} {\bibfnamefont {T.}~\bibnamefont {Pfau}},\ }\href
  {http://stacks.iop.org/0034-4885/72/i=12/a=126401} {\bibfield  {journal}
  {\bibinfo  {journal} {Rep. Prog. Phys.}\ }\textbf {\bibinfo {volume} {72}},\
  \bibinfo {pages} {126401} (\bibinfo {year} {2009}{\natexlab{a}})}\BibitemShut
  {NoStop}%
\bibitem [{\citenamefont {Ni}\ \emph {et~al.}(2008)\citenamefont {Ni},
  \citenamefont {Ospelkaus}, \citenamefont {de~Miranda}, \citenamefont {Pe'er},
  \citenamefont {Neyenhuis}, \citenamefont {Zirbel}, \citenamefont
  {Kotochigova}, \citenamefont {Julienne}, \citenamefont {Jin},\ and\
  \citenamefont {Ye}}]{Ni2008}%
  \BibitemOpen
  \bibfield  {author} {\bibinfo {author} {\bibfnamefont {K.-K.}\ \bibnamefont
  {Ni}}, \bibinfo {author} {\bibfnamefont {S.}~\bibnamefont {Ospelkaus}},
  \bibinfo {author} {\bibfnamefont {M.~H.~G.}\ \bibnamefont {de~Miranda}},
  \bibinfo {author} {\bibfnamefont {A.}~\bibnamefont {Pe'er}}, \bibinfo
  {author} {\bibfnamefont {B.}~\bibnamefont {Neyenhuis}}, \bibinfo {author}
  {\bibfnamefont {J.~J.}\ \bibnamefont {Zirbel}}, \bibinfo {author}
  {\bibfnamefont {S.}~\bibnamefont {Kotochigova}}, \bibinfo {author}
  {\bibfnamefont {P.~S.}\ \bibnamefont {Julienne}}, \bibinfo {author}
  {\bibfnamefont {D.~S.}\ \bibnamefont {Jin}}, \ and\ \bibinfo {author}
  {\bibfnamefont {J.}~\bibnamefont {Ye}},\ }\href {\doibase
  10.1126/science.1163861} {\bibfield  {journal} {\bibinfo  {journal}
  {Science}\ }\textbf {\bibinfo {volume} {322}},\ \bibinfo {pages} {231}
  (\bibinfo {year} {2008})}\BibitemShut {NoStop}%
\bibitem [{\citenamefont {Deiglmayr}\ \emph {et~al.}(2008)\citenamefont
  {Deiglmayr}, \citenamefont {Grochola}, \citenamefont {Repp}, \citenamefont
  {M\"ortlbauer}, \citenamefont {Gl\"uck}, \citenamefont {Lange}, \citenamefont
  {Dulieu}, \citenamefont {Wester},\ and\ \citenamefont
  {Weidem\"uller}}]{Deiglmayr2008}%
  \BibitemOpen
  \bibfield  {author} {\bibinfo {author} {\bibfnamefont {J.}~\bibnamefont
  {Deiglmayr}}, \bibinfo {author} {\bibfnamefont {A.}~\bibnamefont {Grochola}},
  \bibinfo {author} {\bibfnamefont {M.}~\bibnamefont {Repp}}, \bibinfo {author}
  {\bibfnamefont {K.}~\bibnamefont {M\"ortlbauer}}, \bibinfo {author}
  {\bibfnamefont {C.}~\bibnamefont {Gl\"uck}}, \bibinfo {author} {\bibfnamefont
  {J.}~\bibnamefont {Lange}}, \bibinfo {author} {\bibfnamefont
  {O.}~\bibnamefont {Dulieu}}, \bibinfo {author} {\bibfnamefont
  {R.}~\bibnamefont {Wester}}, \ and\ \bibinfo {author} {\bibfnamefont
  {M.}~\bibnamefont {Weidem\"uller}},\ }\href {\doibase
  10.1103/PhysRevLett.101.133004} {\bibfield  {journal} {\bibinfo  {journal}
  {Phys. Rev. Lett.}\ }\textbf {\bibinfo {volume} {101}},\ \bibinfo {pages}
  {133004} (\bibinfo {year} {2008})}\BibitemShut {NoStop}%
\bibitem [{\citenamefont {Stuhl}\ \emph {et~al.}(2012)\citenamefont {Stuhl},
  \citenamefont {Hummon}, \citenamefont {Yeo}, \citenamefont {Quemener},
  \citenamefont {Bohn},\ and\ \citenamefont {Ye}}]{Stuhl2012}%
  \BibitemOpen
  \bibfield  {author} {\bibinfo {author} {\bibfnamefont {B.~K.}\ \bibnamefont
  {Stuhl}}, \bibinfo {author} {\bibfnamefont {M.~T.}\ \bibnamefont {Hummon}},
  \bibinfo {author} {\bibfnamefont {M.}~\bibnamefont {Yeo}}, \bibinfo {author}
  {\bibfnamefont {G.}~\bibnamefont {Quemener}}, \bibinfo {author}
  {\bibfnamefont {J.~L.}\ \bibnamefont {Bohn}}, \ and\ \bibinfo {author}
  {\bibfnamefont {J.}~\bibnamefont {Ye}},\ }\href {\doibase
  10.1038/nature11718} {\bibfield  {journal} {\bibinfo  {journal} {Nature}\
  }\textbf {\bibinfo {volume} {492}},\ \bibinfo {pages} {396} (\bibinfo {year}
  {2012})}\BibitemShut {NoStop}%
\bibitem [{\citenamefont {Menotti}\ \emph {et~al.}(2008)\citenamefont
  {Menotti}, \citenamefont {Lewenstein}, \citenamefont {Lahaye},\ and\
  \citenamefont {Pfau}}]{Menotti2008}%
  \BibitemOpen
  \bibfield  {author} {\bibinfo {author} {\bibfnamefont {C.}~\bibnamefont
  {Menotti}}, \bibinfo {author} {\bibfnamefont {M.}~\bibnamefont {Lewenstein}},
  \bibinfo {author} {\bibfnamefont {T.}~\bibnamefont {Lahaye}}, \ and\ \bibinfo
  {author} {\bibfnamefont {T.}~\bibnamefont {Pfau}},\ }\href {\doibase
  10.1063/1.2839130} {\bibfield  {journal} {\bibinfo  {journal} {AIP Conference
  Proceedings}\ }\textbf {\bibinfo {volume} {970}},\ \bibinfo {pages} {332}
  (\bibinfo {year} {2008})}\BibitemShut {NoStop}%
\bibitem [{\citenamefont {Koch}\ \emph {et~al.}(2008)\citenamefont {Koch},
  \citenamefont {Lahaye}, \citenamefont {Metz}, \citenamefont {Frohlich},
  \citenamefont {Griesmaier},\ and\ \citenamefont {Pfau}}]{Koch2008}%
  \BibitemOpen
  \bibfield  {author} {\bibinfo {author} {\bibfnamefont {T.}~\bibnamefont
  {Koch}}, \bibinfo {author} {\bibfnamefont {T.}~\bibnamefont {Lahaye}},
  \bibinfo {author} {\bibfnamefont {J.}~\bibnamefont {Metz}}, \bibinfo {author}
  {\bibfnamefont {B.}~\bibnamefont {Frohlich}}, \bibinfo {author}
  {\bibfnamefont {A.}~\bibnamefont {Griesmaier}}, \ and\ \bibinfo {author}
  {\bibfnamefont {T.}~\bibnamefont {Pfau}},\ }\href
  {http://dx.doi.org/10.1038/nphys887} {\bibfield  {journal} {\bibinfo
  {journal} {Nat. Phys.}\ }\textbf {\bibinfo {volume} {4}},\ \bibinfo {pages}
  {218} (\bibinfo {year} {2008})}\BibitemShut {NoStop}%
\bibitem [{\citenamefont {Lahaye}\ \emph
  {et~al.}(2009{\natexlab{b}})\citenamefont {Lahaye}, \citenamefont {Metz},
  \citenamefont {Koch}, \citenamefont {Fröhlich}, \citenamefont {Griesmaier},\
  and\ \citenamefont {Pfau}}]{Lahaye2009b}%
  \BibitemOpen
  \bibfield  {author} {\bibinfo {author} {\bibfnamefont {T.}~\bibnamefont
  {Lahaye}}, \bibinfo {author} {\bibfnamefont {J.}~\bibnamefont {Metz}},
  \bibinfo {author} {\bibfnamefont {T.}~\bibnamefont {Koch}}, \bibinfo {author}
  {\bibfnamefont {B.}~\bibnamefont {Fröhlich}}, \bibinfo {author}
  {\bibfnamefont {A.~.}\ \bibnamefont {Griesmaier}}, \ and\ \bibinfo {author}
  {\bibfnamefont {T.}~\bibnamefont {Pfau}},\ }in\ \href@noop {} {\emph
  {\bibinfo {booktitle} {Pushing The Frontiers Of Atomic Physics: Proc. 21st
  Int. Conf. on Atomic Physics}}},\ \bibinfo {editor} {edited by\ \bibinfo
  {editor} {\bibfnamefont {R.}~\bibnamefont {C\^ot\'e}}}\ (\bibinfo
  {publisher} {World Scientific},\ \bibinfo {year} {2009})\BibitemShut
  {NoStop}%
\bibitem [{\citenamefont {Micheli}\ \emph {et~al.}(2006)\citenamefont
  {Micheli}, \citenamefont {Brennen},\ and\ \citenamefont
  {Zoller}}]{Micheli2006}%
  \BibitemOpen
  \bibfield  {author} {\bibinfo {author} {\bibfnamefont {A.}~\bibnamefont
  {Micheli}}, \bibinfo {author} {\bibfnamefont {G.~K.}\ \bibnamefont
  {Brennen}}, \ and\ \bibinfo {author} {\bibfnamefont {P.}~\bibnamefont
  {Zoller}},\ }\href {http://dx.doi.org/10.1038/nphys287} {\bibfield  {journal}
  {\bibinfo  {journal} {Nat. Phys.}\ }\textbf {\bibinfo {volume} {2}},\
  \bibinfo {pages} {341} (\bibinfo {year} {2006})}\BibitemShut {NoStop}%
\bibitem [{\citenamefont {Baranov}\ \emph {et~al.}(2012)\citenamefont
  {Baranov}, \citenamefont {Dalmonte}, \citenamefont {Pupillo},\ and\
  \citenamefont {Zoller}}]{Baranov2012}%
  \BibitemOpen
  \bibfield  {author} {\bibinfo {author} {\bibfnamefont {M.~A.}\ \bibnamefont
  {Baranov}}, \bibinfo {author} {\bibfnamefont {M.}~\bibnamefont {Dalmonte}},
  \bibinfo {author} {\bibfnamefont {G.}~\bibnamefont {Pupillo}}, \ and\
  \bibinfo {author} {\bibfnamefont {P.}~\bibnamefont {Zoller}},\ }\href
  {\doibase 10.1021/cr2003568} {\bibfield  {journal} {\bibinfo  {journal}
  {Chem. Rev.}\ }\textbf {\bibinfo {volume} {112}},\ \bibinfo {pages} {5012}
  (\bibinfo {year} {2012})}\BibitemShut {NoStop}%
\bibitem [{\citenamefont {Ospelkaus}\ \emph {et~al.}(2010)\citenamefont
  {Ospelkaus}, \citenamefont {Ni}, \citenamefont {Wang}, \citenamefont
  {de~Miranda}, \citenamefont {Neyenhuis}, \citenamefont {Quéméner},
  \citenamefont {Julienne}, \citenamefont {Bohn}, \citenamefont {Jin},\ and\
  \citenamefont {Ye}}]{Ospelkaus2010}%
  \BibitemOpen
  \bibfield  {author} {\bibinfo {author} {\bibfnamefont {S.}~\bibnamefont
  {Ospelkaus}}, \bibinfo {author} {\bibfnamefont {K.-K.}\ \bibnamefont {Ni}},
  \bibinfo {author} {\bibfnamefont {D.}~\bibnamefont {Wang}}, \bibinfo {author}
  {\bibfnamefont {M.~H.~G.}\ \bibnamefont {de~Miranda}}, \bibinfo {author}
  {\bibfnamefont {B.}~\bibnamefont {Neyenhuis}}, \bibinfo {author}
  {\bibfnamefont {G.}~\bibnamefont {Quéméner}}, \bibinfo {author}
  {\bibfnamefont {P.~S.}\ \bibnamefont {Julienne}}, \bibinfo {author}
  {\bibfnamefont {J.~L.}\ \bibnamefont {Bohn}}, \bibinfo {author}
  {\bibfnamefont {D.~S.}\ \bibnamefont {Jin}}, \ and\ \bibinfo {author}
  {\bibfnamefont {J.}~\bibnamefont {Ye}},\ }\href {\doibase
  10.1126/science.1184121} {\bibfield  {journal} {\bibinfo  {journal}
  {Science}\ }\textbf {\bibinfo {volume} {327}},\ \bibinfo {pages} {853}
  (\bibinfo {year} {2010})}\BibitemShut {NoStop}%
\bibitem [{\citenamefont {Qu\'em\'ener}\ and\ \citenamefont
  {Julienne}(2012)}]{Quemener2012}%
  \BibitemOpen
  \bibfield  {author} {\bibinfo {author} {\bibfnamefont {G.}~\bibnamefont
  {Qu\'em\'ener}}\ and\ \bibinfo {author} {\bibfnamefont {P.~S.}\ \bibnamefont
  {Julienne}},\ }\href {\doibase 10.1021/cr300092g} {\bibfield  {journal}
  {\bibinfo  {journal} {Chem. Rev.}\ }\textbf {\bibinfo {volume} {112}},\
  \bibinfo {pages} {4949} (\bibinfo {year} {2012})}\BibitemShut {NoStop}%
\bibitem [{\citenamefont {Deb}\ and\ \citenamefont {You}(2001)}]{Deb2001}%
  \BibitemOpen
  \bibfield  {author} {\bibinfo {author} {\bibfnamefont {B.}~\bibnamefont
  {Deb}}\ and\ \bibinfo {author} {\bibfnamefont {L.}~\bibnamefont {You}},\
  }\href {\doibase 10.1103/PhysRevA.64.022717} {\bibfield  {journal} {\bibinfo
  {journal} {Phys. Rev. A}\ }\textbf {\bibinfo {volume} {64}},\ \bibinfo
  {pages} {022717} (\bibinfo {year} {2001})}\BibitemShut {NoStop}%
\bibitem [{\citenamefont {Gao}(1999)}]{Gao1999}%
  \BibitemOpen
  \bibfield  {author} {\bibinfo {author} {\bibfnamefont {B.}~\bibnamefont
  {Gao}},\ }\href {\doibase 10.1103/PhysRevA.59.2778} {\bibfield  {journal}
  {\bibinfo  {journal} {Phys. Rev. A}\ }\textbf {\bibinfo {volume} {59}},\
  \bibinfo {pages} {2778} (\bibinfo {year} {1999})}\BibitemShut {NoStop}%
\bibitem [{\citenamefont {Kanjilal}\ and\ \citenamefont
  {Blume}(2008)}]{Kanjilal2008}%
  \BibitemOpen
  \bibfield  {author} {\bibinfo {author} {\bibfnamefont {K.}~\bibnamefont
  {Kanjilal}}\ and\ \bibinfo {author} {\bibfnamefont {D.}~\bibnamefont
  {Blume}},\ }\href {\doibase 10.1103/PhysRevA.78.040703} {\bibfield  {journal}
  {\bibinfo  {journal} {Phys. Rev. A}\ }\textbf {\bibinfo {volume} {78}},\
  \bibinfo {pages} {040703} (\bibinfo {year} {2008})}\BibitemShut {NoStop}%
\bibitem [{\citenamefont {Melezhik}\ and\ \citenamefont
  {Hu}(2003)}]{Melezhik2003}%
  \BibitemOpen
  \bibfield  {author} {\bibinfo {author} {\bibfnamefont {V.~S.}\ \bibnamefont
  {Melezhik}}\ and\ \bibinfo {author} {\bibfnamefont {C.-Y.}\ \bibnamefont
  {Hu}},\ }\href {\doibase 10.1103/PhysRevLett.90.083202} {\bibfield  {journal}
  {\bibinfo  {journal} {Phys. Rev. Lett.}\ }\textbf {\bibinfo {volume} {90}},\
  \bibinfo {pages} {083202} (\bibinfo {year} {2003})}\BibitemShut {NoStop}%
\bibitem [{\citenamefont {Derevianko}(2003)}]{Derevianko2003}%
  \BibitemOpen
  \bibfield  {author} {\bibinfo {author} {\bibfnamefont {A.}~\bibnamefont
  {Derevianko}},\ }\href {\doibase 10.1103/PhysRevA.67.033607} {\bibfield
  {journal} {\bibinfo  {journal} {Phys. Rev. A}\ }\textbf {\bibinfo {volume}
  {67}},\ \bibinfo {pages} {033607} (\bibinfo {year} {2003})}\BibitemShut
  {NoStop}%
\bibitem [{\citenamefont {Wang}\ \emph
  {et~al.}(2011{\natexlab{a}})\citenamefont {Wang}, \citenamefont {D'Incao},\
  and\ \citenamefont {Greene}}]{Wang2011}%
  \BibitemOpen
  \bibfield  {author} {\bibinfo {author} {\bibfnamefont {Y.}~\bibnamefont
  {Wang}}, \bibinfo {author} {\bibfnamefont {J.~P.}\ \bibnamefont {D'Incao}}, \
  and\ \bibinfo {author} {\bibfnamefont {C.~H.}\ \bibnamefont {Greene}},\
  }\href {\doibase 10.1103/PhysRevLett.106.233201} {\bibfield  {journal}
  {\bibinfo  {journal} {Phys. Rev. Lett.}\ }\textbf {\bibinfo {volume} {106}},\
  \bibinfo {pages} {233201} (\bibinfo {year} {2011}{\natexlab{a}})}\BibitemShut
  {NoStop}%
\bibitem [{\citenamefont {Wang}\ \emph
  {et~al.}(2011{\natexlab{b}})\citenamefont {Wang}, \citenamefont {D'Incao},\
  and\ \citenamefont {Greene}}]{Wang2011b}%
  \BibitemOpen
  \bibfield  {author} {\bibinfo {author} {\bibfnamefont {Y.}~\bibnamefont
  {Wang}}, \bibinfo {author} {\bibfnamefont {J.~P.}\ \bibnamefont {D'Incao}}, \
  and\ \bibinfo {author} {\bibfnamefont {C.~H.}\ \bibnamefont {Greene}},\
  }\href {\doibase 10.1103/PhysRevLett.107.233201} {\bibfield  {journal}
  {\bibinfo  {journal} {Phys. Rev. Lett.}\ }\textbf {\bibinfo {volume} {107}},\
  \bibinfo {pages} {233201} (\bibinfo {year} {2011}{\natexlab{b}})}\BibitemShut
  {NoStop}%
\bibitem [{\citenamefont {Volosniev}\ \emph {et~al.}(2013)\citenamefont
  {Volosniev}, \citenamefont {Armstrong}, \citenamefont {Fedorov},
  \citenamefont {Jensen}, \citenamefont {Valiente},\ and\ \citenamefont
  {Zinner}}]{Volosniev2013}%
  \BibitemOpen
  \bibfield  {author} {\bibinfo {author} {\bibfnamefont {A.~G.}\ \bibnamefont
  {Volosniev}}, \bibinfo {author} {\bibfnamefont {J.~R.}\ \bibnamefont
  {Armstrong}}, \bibinfo {author} {\bibfnamefont {D.~V.}\ \bibnamefont
  {Fedorov}}, \bibinfo {author} {\bibfnamefont {A.~S.}\ \bibnamefont {Jensen}},
  \bibinfo {author} {\bibfnamefont {M.}~\bibnamefont {Valiente}}, \ and\
  \bibinfo {author} {\bibfnamefont {N.~T.}\ \bibnamefont {Zinner}},\ }\href
  {http://stacks.iop.org/1367-2630/15/i=4/a=043046} {\bibfield  {journal}
  {\bibinfo  {journal} {New J. Phys.}\ }\textbf {\bibinfo {volume} {15}},\
  \bibinfo {pages} {043046} (\bibinfo {year} {2013})}\BibitemShut {NoStop}%
\bibitem [{\citenamefont {Ticknor}(2009)}]{Ticknor2009}%
  \BibitemOpen
  \bibfield  {author} {\bibinfo {author} {\bibfnamefont {C.}~\bibnamefont
  {Ticknor}},\ }\href {\doibase 10.1103/PhysRevA.80.052702} {\bibfield
  {journal} {\bibinfo  {journal} {Phys. Rev. A}\ }\textbf {\bibinfo {volume}
  {80}},\ \bibinfo {pages} {052702} (\bibinfo {year} {2009})}\BibitemShut
  {NoStop}%
\bibitem [{\citenamefont {Ticknor}(2010)}]{Ticknor2010}%
  \BibitemOpen
  \bibfield  {author} {\bibinfo {author} {\bibfnamefont {C.}~\bibnamefont
  {Ticknor}},\ }\href {\doibase 10.1103/PhysRevA.81.042708} {\bibfield
  {journal} {\bibinfo  {journal} {Phys. Rev. A}\ }\textbf {\bibinfo {volume}
  {81}},\ \bibinfo {pages} {042708} (\bibinfo {year} {2010})}\BibitemShut
  {NoStop}%
\bibitem [{\citenamefont {D'Incao}\ and\ \citenamefont
  {Greene}(2011)}]{DIncao2011}%
  \BibitemOpen
  \bibfield  {author} {\bibinfo {author} {\bibfnamefont {J.~P.}\ \bibnamefont
  {D'Incao}}\ and\ \bibinfo {author} {\bibfnamefont {C.~H.}\ \bibnamefont
  {Greene}},\ }\href {\doibase 10.1103/PhysRevA.83.030702} {\bibfield
  {journal} {\bibinfo  {journal} {Phys. Rev. A}\ }\textbf {\bibinfo {volume}
  {83}},\ \bibinfo {pages} {030702} (\bibinfo {year} {2011})}\BibitemShut
  {NoStop}%
\bibitem [{\citenamefont {Cai}\ \emph {et~al.}(2010)\citenamefont {Cai},
  \citenamefont {Rosenkranz}, \citenamefont {Lei},\ and\ \citenamefont
  {Bao}}]{Cai2010}%
  \BibitemOpen
  \bibfield  {author} {\bibinfo {author} {\bibfnamefont {Y.}~\bibnamefont
  {Cai}}, \bibinfo {author} {\bibfnamefont {M.}~\bibnamefont {Rosenkranz}},
  \bibinfo {author} {\bibfnamefont {Z.}~\bibnamefont {Lei}}, \ and\ \bibinfo
  {author} {\bibfnamefont {W.}~\bibnamefont {Bao}},\ }\href {\doibase
  10.1103/PhysRevA.82.043623} {\bibfield  {journal} {\bibinfo  {journal} {Phys.
  Rev. A}\ }\textbf {\bibinfo {volume} {82}},\ \bibinfo {pages} {043623}
  (\bibinfo {year} {2010})}\BibitemShut {NoStop}%
\bibitem [{\citenamefont {Koval}\ \emph {et~al.}(2014)\citenamefont {Koval},
  \citenamefont {Koval},\ and\ \citenamefont {Melezhik}}]{Koval2014}%
  \BibitemOpen
  \bibfield  {author} {\bibinfo {author} {\bibfnamefont {E.~A.}\ \bibnamefont
  {Koval}}, \bibinfo {author} {\bibfnamefont {O.~A.}\ \bibnamefont {Koval}}, \
  and\ \bibinfo {author} {\bibfnamefont {V.~S.}\ \bibnamefont {Melezhik}},\
  }\href {\doibase 10.1103/PhysRevA.89.052710} {\bibfield  {journal} {\bibinfo
  {journal} {Phys. Rev. A}\ }\textbf {\bibinfo {volume} {89}},\ \bibinfo
  {pages} {052710} (\bibinfo {year} {2014})}\BibitemShut {NoStop}%
\bibitem [{\citenamefont {Volosniev}\ \emph {et~al.}(2011)\citenamefont
  {Volosniev}, \citenamefont {Zinner}, \citenamefont {Fedorov}, \citenamefont
  {Jensen},\ and\ \citenamefont {Wunsch}}]{Volosniev2011}%
  \BibitemOpen
  \bibfield  {author} {\bibinfo {author} {\bibfnamefont {A.~G.}\ \bibnamefont
  {Volosniev}}, \bibinfo {author} {\bibfnamefont {N.~T.}\ \bibnamefont
  {Zinner}}, \bibinfo {author} {\bibfnamefont {D.~V.}\ \bibnamefont {Fedorov}},
  \bibinfo {author} {\bibfnamefont {A.~S.}\ \bibnamefont {Jensen}}, \ and\
  \bibinfo {author} {\bibfnamefont {B.}~\bibnamefont {Wunsch}},\ }\href
  {http://stacks.iop.org/0953-4075/44/i=12/a=125301} {\bibfield  {journal}
  {\bibinfo  {journal} {J. Phys. B: At., Mol. Opt. Phys.}\ }\textbf {\bibinfo
  {volume} {44}},\ \bibinfo {pages} {125301} (\bibinfo {year}
  {2011})}\BibitemShut {NoStop}%
\bibitem [{\citenamefont {Giannakeas}\ \emph {et~al.}(2013)\citenamefont
  {Giannakeas}, \citenamefont {Melezhik},\ and\ \citenamefont
  {Schmelcher}}]{Giannakeas2013}%
  \BibitemOpen
  \bibfield  {author} {\bibinfo {author} {\bibfnamefont {P.}~\bibnamefont
  {Giannakeas}}, \bibinfo {author} {\bibfnamefont {V.~S.}\ \bibnamefont
  {Melezhik}}, \ and\ \bibinfo {author} {\bibfnamefont {P.}~\bibnamefont
  {Schmelcher}},\ }\href {\doibase 10.1103/PhysRevLett.111.183201} {\bibfield
  {journal} {\bibinfo  {journal} {Phys. Rev. Lett.}\ }\textbf {\bibinfo
  {volume} {111}},\ \bibinfo {pages} {183201} (\bibinfo {year}
  {2013})}\BibitemShut {NoStop}%
\bibitem [{\citenamefont {Kanjilal}\ \emph {et~al.}(2007)\citenamefont
  {Kanjilal}, \citenamefont {Bohn},\ and\ \citenamefont
  {Blume}}]{Kanjilal2007}%
  \BibitemOpen
  \bibfield  {author} {\bibinfo {author} {\bibfnamefont {K.}~\bibnamefont
  {Kanjilal}}, \bibinfo {author} {\bibfnamefont {J.}~\bibnamefont {Bohn}}, \
  and\ \bibinfo {author} {\bibfnamefont {D.}~\bibnamefont {Blume}},\ }\href
  {\doibase 10.1103/PhysRevA.75.052705} {\bibfield  {journal} {\bibinfo
  {journal} {Phys. Rev. A}\ }\textbf {\bibinfo {volume} {75}},\ \bibinfo
  {pages} {052705} (\bibinfo {year} {2007})}\BibitemShut {NoStop}%
\bibitem [{\citenamefont {Wang}\ and\ \citenamefont {Greene}(2012)}]{Wang2012}%
  \BibitemOpen
  \bibfield  {author} {\bibinfo {author} {\bibfnamefont {Y.}~\bibnamefont
  {Wang}}\ and\ \bibinfo {author} {\bibfnamefont {C.~H.}\ \bibnamefont
  {Greene}},\ }\href {\doibase 10.1103/PhysRevA.85.022704} {\bibfield
  {journal} {\bibinfo  {journal} {Phys. Rev. A}\ }\textbf {\bibinfo {volume}
  {85}},\ \bibinfo {pages} {022704} (\bibinfo {year} {2012})}\BibitemShut
  {NoStop}%
\bibitem [{\citenamefont {Bradly}\ \emph {et~al.}(2014)\citenamefont {Bradly},
  \citenamefont {Mulkerin}, \citenamefont {Martin},\ and\ \citenamefont
  {Quiney}}]{Bradly2014}%
  \BibitemOpen
  \bibfield  {author} {\bibinfo {author} {\bibfnamefont {C.~J.}\ \bibnamefont
  {Bradly}}, \bibinfo {author} {\bibfnamefont {B.~C.}\ \bibnamefont
  {Mulkerin}}, \bibinfo {author} {\bibfnamefont {A.~M.}\ \bibnamefont
  {Martin}}, \ and\ \bibinfo {author} {\bibfnamefont {H.~M.}\ \bibnamefont
  {Quiney}},\ }\href {\doibase 10.1103/PhysRevA.90.023626} {\bibfield
  {journal} {\bibinfo  {journal} {Phys. Rev. A}\ }\textbf {\bibinfo {volume}
  {90}},\ \bibinfo {pages} {023626} (\bibinfo {year} {2014})}\BibitemShut
  {NoStop}%
\bibitem [{\citenamefont {Suzuki}\ and\ \citenamefont
  {Varga}(1998)}]{Suzuki1998}%
  \BibitemOpen
  \bibfield  {author} {\bibinfo {author} {\bibfnamefont {Y.}~\bibnamefont
  {Suzuki}}\ and\ \bibinfo {author} {\bibfnamefont {K.}~\bibnamefont {Varga}},\
  }\href@noop {} {\emph {\bibinfo {title} {Stochastic Variational Approach to
  Quantum-Mechanical Few-Body Problems}}}\ (\bibinfo  {publisher}
  {Springer-Verlag},\ \bibinfo {year} {1998})\BibitemShut {NoStop}%
\bibitem [{\citenamefont {Mitroy}\ \emph {et~al.}(2013)\citenamefont {Mitroy},
  \citenamefont {Bubin}, \citenamefont {Horiuchi}, \citenamefont {Suzuki},
  \citenamefont {Adamowicz}, \citenamefont {Cencek}, \citenamefont {Szalewicz},
  \citenamefont {Komasa}, \citenamefont {Blume},\ and\ \citenamefont
  {Varga}}]{Mitroy2013}%
  \BibitemOpen
  \bibfield  {author} {\bibinfo {author} {\bibfnamefont {J.}~\bibnamefont
  {Mitroy}}, \bibinfo {author} {\bibfnamefont {S.}~\bibnamefont {Bubin}},
  \bibinfo {author} {\bibfnamefont {W.}~\bibnamefont {Horiuchi}}, \bibinfo
  {author} {\bibfnamefont {Y.}~\bibnamefont {Suzuki}}, \bibinfo {author}
  {\bibfnamefont {L.}~\bibnamefont {Adamowicz}}, \bibinfo {author}
  {\bibfnamefont {W.}~\bibnamefont {Cencek}}, \bibinfo {author} {\bibfnamefont
  {K.}~\bibnamefont {Szalewicz}}, \bibinfo {author} {\bibfnamefont
  {J.}~\bibnamefont {Komasa}}, \bibinfo {author} {\bibfnamefont
  {D.}~\bibnamefont {Blume}}, \ and\ \bibinfo {author} {\bibfnamefont
  {K.}~\bibnamefont {Varga}},\ }\href {\doibase 10.1103/RevModPhys.85.693}
  {\bibfield  {journal} {\bibinfo  {journal} {Rev. Mod. Phys.}\ }\textbf
  {\bibinfo {volume} {85}},\ \bibinfo {pages} {693} (\bibinfo {year}
  {2013})}\BibitemShut {NoStop}%
\bibitem [{\citenamefont {Johnson}(1973)}]{Johnson1973}%
  \BibitemOpen
  \bibfield  {author} {\bibinfo {author} {\bibfnamefont {B.}~\bibnamefont
  {Johnson}},\ }\href {\doibase http://dx.doi.org/10.1016/0021-9991(73)90049-1}
  {\bibfield  {journal} {\bibinfo  {journal} {J. Comput. Phys.}\ }\textbf
  {\bibinfo {volume} {13}},\ \bibinfo {pages} {445 } (\bibinfo {year}
  {1973})}\BibitemShut {NoStop}%
\end{thebibliography}%

\end{document}